\documentstyle[11pt,aaspp4]{article}

\begin{document}
\newcommand{\etal}{et al. }
\newcommand{\tsph}{TreeSPH}
\newcommand{\K}{{\rm K}}

\newbox\grsign \setbox\grsign=\hbox{$>$} \newdimen\grdimen \grdimen=\ht\grsign
\newbox\simlessbox \newbox\simgreatbox
\setbox\simgreatbox=\hbox{\raise.5ex\hbox{$>$}\llap
     {\lower.5ex\hbox{$\sim$}}}\ht1=\grdimen\dp1=0pt
\setbox\simlessbox=\hbox{\raise.5ex\hbox{$<$}\llap
     {\lower.5ex\hbox{$\sim$}}}\ht2=\grdimen\dp2=0pt
\newcommand{\simgt}{\mathrel{\copy\simgreatbox}}
\newcommand{\simlt}{\mathrel{\copy\simlessbox}}
\newcommand\junits{{\rm erg\,s}^{-1}\,{\rm cm}^{-2}\,{\rm sr}^{-1}\,
		   {\rm Hz}^{-1}}
\newcommand\iunits{{\rm cm}^{3}\,{\rm s}^{-1}}

\slugcomment{Submitted to ApJ}

\lefthead{Hellsten et. al.}
\righthead{Metal Lines Associated with Lyman Alpha Absorbers}

\title{Metal Lines Associated with Lyman Alpha Absorbers: A Comparison
       of Theory and Observations}

\author{Uffe Hellsten, Romeel Dav\'{e}, and Lars Hernquist\altaffilmark{1}}
\affil{University of California, Lick Observatory, Santa Cruz, CA 95064}
\author{David H. Weinberg}
\affil{Ohio State University, Department of Astronomy, Columbus, OH 43210} 
\and
\author{Neal Katz}
\affil{University of Washington, Department of Astronomy, Seattle, WA 98195}
\altaffiltext{1}{Presidential Faculty Fellow}

\begin{abstract}
We study metal line absorption of CIV, CII, SiIV, and NV  
at redshifts $z=3.5$ to $z=2$ within the framework of a cosmological model 
for the Lyman alpha forest, comparing the results of numerical simulations
to recent observations by Songaila \& Cowie (1996, SC).  
In agreement with Rauch, Haehnelt \& Steinmetz (1996), we find
that the observed mean value of the CIV/HI ratio at $z \simeq 3$ is 
reproduced if a uniform metallicity of $[{\rm C}/{\rm H}]\sim -2.5$ is assumed
in our model,
but that the observed scatter in this ratio is larger than predicted, 
implying a scatter in the metallicity of the absorbing systems of roughly an
order of magnitude. 
We further argue that absorbers with relatively low column densities
($\log{N_{\rm HI}} < 15$) likely have a mean metallicity [C/H] less 
than $-2.5$,  a result which is basically independent of the  
model considered. The enrichment pattern that is required for our 
model to match SC's CIV observations is very similar to that {\it predicted}
by Gnedin \& Ostriker's (1997) simulations of reionization and metal
enrichment by population III stars in this type of cosmological scenario.

Our model predicts no significant evolution in the mean values of metal
line column densities between $z=3.5$ and $z=2$.
Comparison of the predicted and observed numbers of 
SiIV and NV systems suggests that the photoionizing background radiation
field at $z\sim 3$ is somewhat softer than that proposed by Haardt \& Madau 
(1996).
Our model does not account for the increase
in the SiIV/CIV ratio at $z \simgt 3.2$ found by SC. 
While SC suggested that the increase could be 
explained by a softening of the
radiation field due to HeII absorption at $z \simgt 3$, 
such a modification does not significantly raise the mean value
of SiIV/CIV in our simulation
because it shifts numerous weak SiIV systems 
to just above the detection limit, thus keeping the mean column density
of observable SiIV systems low. 
\end{abstract}


\section{Introduction}
Spectra of high redshift quasi-stellar objects (QSOs) are densely populated
with absorption lines blueward of the Lyman alpha emission peak. This 
so-called ``Lyman alpha forest'' is interpreted as arising from
absorption by neutral hydrogen cosmologically distributed
along the line-of-sight to the QSO (e.g. Lynds 1971;
Sargent \etal 1980). For more than a decade it was not known whether
the gaseous systems responsible for the Lyman alpha forest were  
chemically pristine. Meyer \& York (1987) presented early evidence for metal 
lines in the forest, and within the last several years observations with
improved spectral resolution and signal-to-noise ratio (S/N) have revealed the 
presence of heavy elements in a significant fraction of Lyman alpha
systems with moderate HI column densities ($\sim 10^{14}-10^{17} 
{\rm cm}^{-2}$). Analyzing data from the HIRES spectrograph on the Keck 10~m 
telescope, Cowie \etal (1995) find that $\sim 50$\%
of identified Lyman alpha lines with $\log{{\rm N}_{\rm HI}} > 14.5 $ have
associated CIV absorption. This finding is roughly
consistent with that of Womble \etal (1995) who infer CIV
absorption in 40--45\% of all Lyman alpha systems with 
$\log{{\rm N}_{\rm HI}} > 14.3$.
More recently, Songaila \& Cowie (1996, hereafter SC) presented an analysis of 
even higher S/N spectra and found CIV absorption in $\simeq 75\%$ of
all systems with $\log{{\rm N}_{\rm HI}} > 14.5$ and in $\simeq 90-100$ \% of 
those with $\log{{\rm N}_{\rm HI}} > 15.2$. In addition, they detected 
absorption from SiIV, CII, and NV in a number of these systems.  

In the present paper, we analyze the predicted metal line absorption from
a hydrodynamic cosmological simulation of the high-redshift intergalactic
medium, and compare these predictions to the observations of SC.
The analysis of such simulations during recent years (e.g. Cen \etal 1994; 
Zhang \etal 1995; Petitjean \etal 1995; Hernquist \etal 1996) has 
established a convincing general scenario in which the Lyman alpha forest 
is produced by regions of moderate overdensity in a hierarchically clustering 
universe, pervaded by an
ionizing radiation field presumed to originate from 
the QSOs (for related semi-analytic modeling see, e.g., Bi 1993; Bi \&
Davidsen 1996; Hui, Gnedin, \& Zhang 1997). These cosmological Lyman alpha 
forest models predict QSO absorption features that resemble those observed, 
and they reproduce the column density and 
Doppler parameter distributions down to the lowest observable column densities 
(Miralda-Escud\'e \etal 1996; 
Dav\'e \etal 1997, hereafter DHWK; Zhang \etal 1997).
The interpretation of the metal line data within the 
framework of these models of the Lyman alpha forest constitutes a further test
of the validity of the models and provides potentially valuable insight to the 
physical conditions prevailing in the intergalactic medium, such as gas 
densities, temperatures, and the shape and intensity of the ionizing radiation
field. It is even conceivable that comparisons between such models and 
the wealth of forthcoming observations will constrain the underlying 
cosmological parameters, although this possibility remains to be demonstrated.

Haehnelt, Steinmetz \& Rauch (1996) and Rauch, Haehnelt \& Steinmetz (1996) 
have pioneered the 
technique for incorporating metals into cosmological Lyman alpha models. 
They endow the gas with a uniform metallicity, postulated to have arisen from a
population III burst of star formation at a significantly earlier time, and then
compute the ionization states based on the assumption of a uniform ionizing 
background of a given strength and spectral shape.
By generating and analyzing artificial metal line spectra along lines of 
sight through cosmological hydrodynamical
simulations, these authors compare their models to 
a number of observables such as the Doppler parameter and column density
distributions, CIV/HI and other metal line ratios, and two point correlation
functions, and generally find acceptable agreement between their models and 
observations. 

In the present work we adopt a similar approach, but whereas  
the gaseous regions studied in the above mentioned works are overdense regions 
which eventually evolve into galaxies, we utilize a larger simulation
which allows us to consider a random region of the universe
and make quantitative predictions regarding the statistics of observable
metal systems and their evolution in redshift.
Through most of this paper, we will assume that
the simulation provides an adequate physical model of the intergalactic
medium and will use the comparison to observations to obtain information about
the metal enrichment of the diffuse gas and the spectral shape of the
ionizing radiation field.  This assumption is supported by
the previous successes of the model mentioned above, 
and we will argue in Section 4 that the success of the model in
predicting the relative numbers of observable metal line systems
provides further evidence favoring this physical picture of the high-redshift
intergalactic medium.

\section{The Model}
The cosmological simulation analyzed here is identical to that described in 
DHWK. A 22.222 comoving Mpc box of baryonic and dark matter is evolved from
$z=49$ to $z=2$ using TreeSPH (Hernquist \& Katz 1989), assuming a CDM 
spectrum of density fluctuations and the parameters 
${\sigma _{8 h^{-1}} = 0.7}$, $h=0.5$, and $\Omega _b = 0.05$, 
where $h \equiv {\rm H}_0/(100 {\rm km/s/Mpc}$).
The mass per gas and dark matter particle is $m_g = 1.5 \times 10^8 M_{\odot}$ 
and $m_{DM} = 2.8 \times 10^9 M_{\odot}$, respectively. 
A spatially uniform photoionizing background 
radiation field is imposed, with the shape suggested by Haardt \& Madau 
(1996; heareafter HM),
and an intensity chosen to reproduce the mean HI optical depth as observed by
Press, Rybicki \& Schneider (1993).
For further details regarding the
simulation method, see Katz, Weinberg \& Hernquist (1996).

Artificial spectra were generated along 400 random lines of sight (LOS) through
the simulation box for each of the redshifts $z$=2.0, 2.33, 2.66, 3.0, and 3.5 
by binning the gas in distance along the LOS and calculating
SPH estimates of density, temperature, and velocity 
$\vec{v}=\vec{v}_{Hubble} + \vec{v}_{pec}$ of the relevant 
species. The mass of a given species, such as CIV, in each SPH particle
is evaluated prior to smoothing, according to 
\begin{equation}
m_{\rm CIV} = m_g \,Z_{\rm C} \,f_{\rm CIV}(\rho,T,J_\nu),
\end{equation}
where $Z_{\rm C}$ is the mass fraction of the metal, here carbon, in
the gas, and $f_{\rm CIV}$ is the fraction of the total carbon atoms
that are found in the ionization stage CIV. Assuming the ionization
to be in or near equilibrium, $f_{\rm CIV}$ is a function of the
local gas density $\rho$, temperature $T$, and the properties of the 
photoionizing background $J_\nu$.  We evaluate $f_{\rm species}$ 
using the code CLOUDY 90 (Ferland 1996), creating a lookup table
in $\rho$ and $T$ given the adopted $J_{\nu}$ and interpolating within this 
table.

The spatial bins are projected into their corresponding velocity bins,
convolving their velocities with a thermal Doppler broadening function.
This procedure yields the redshift-space density profile $n(v)$ 
along the LOS; $n(v)dv/H(z)$ is the column density of a species with
observed radial velocity in the interval $v$ to $v+dv$. 
The optical depth at velocity $v$ is
\begin{equation}
\tau_v = \frac{\pi {\rm e}^2}{m_e c} \, f \lambda H^{-1}(z) n(v), 
\end{equation}
where $f\lambda$ is the product of the oscillator strength and the
rest transition wavelength $\lambda$ for the particular absorption line
(Gunn \& Peterson 1965).
The observed wavelength is $\lambda_o = \lambda \times (1+v/c) \times (1+z)$.

%which results in the column density distribution ${\rm N}_v$, where
%${\rm N}_v dv$ is the column density of a species along the LOS with
%observed radial velocity in the interval $v$ to $v+dv$. 
%The corresponding optical depth $\tau _v$ is given by
%\begin{equation}
%\tau_v = {\rm N}_v \, \frac{\pi {\rm e}^2}{m_e c} \, f \lambda, 
%\end{equation}

The resulting artificial spectra ($e^{-\tau_v}$ vs. $v$) are 
degraded to Keck HIRES resolution ($\Delta\lambda = 0.06$ \AA ),
and noise is added with a S/N comparable to that in the HIRES 
spectrum of Q1422+231 (SC), i.e a S/N of roughly 60 in the forest region and 
upwards of 150 in the region redward of the Lyman alpha emission peak where CIV
is observed.
Voigt profiles are fitted to the resulting artificial spectra using the
automated fitter AUTOVP (DHWK) to obtain the column densities
and $b$-parameters of individual lines.

Given the ionizing background we have employed, we find that
the observed mean metallicity as measured by [CIV/HI] is
best reproduced for a uniform metallicity of [C/H]$=-2.5$.
We will later vary the spectrum of the background to match 
observations of various metals, and when this is done we will
vary [C/H] to match the observed mean column density of CIV.
We further assume [C/Fe]$=0$, [Si/C]$=0.4$, and [N/C]$=-0.7$.  This relative 
abundance pattern is similar to that observed in halo stars, low 
metallicity HII regions, and damped Lyman alpha systems (e.g. Tomkin \etal 1992;
Pettini \etal 1995; Wheeler \etal 1989), although the nitrogen 
abundance is somewhat uncertain.

\section{CIV/HI at $z=3$}
Figure \ref{fig1} shows an example of artificial spectra for Lyman 
alpha (H1216), Lyman beta (H1026), and CIV ($\lambda = 1548\AA$) 
absorption along one of the more spectacular LOS in the sample at $z=3$, fit 
using AUTOVP.  At $v \approx 600$ km/s, a feature saturated 
in both Lyman alpha and Lyman beta is present, possessing a 
column density $N_{\rm HI}= 1.3 \times 10^{16} \, {\rm cm}^{-2}$. 
The corresponding unsaturated CIV 
lines reveal the presence of several velocity components in the system, as is
often observed in high column density Lyman alpha absorbers.
A weaker, more typical absorption feature is seen at 
$v \approx 1300$ km/s. For that system, $N_{\rm HI}=7 \times 10^{14}\, {\rm cm}
^{-2}$, too low to be saturated in Lyman beta, and $N_{\rm CIV}=1.6 
\times 10^{12} \,{\rm cm}^{-2}$, close to the limit of detectability. The 
third system, around $v \approx 1850$ km/s, has $N_{\rm HI}=3 \times 10^{15}\,
{\rm cm}^{-2}$.  This absorber also has several associated CIV lines, and 
distinguishable individual subcomponents in Lyman beta.  When computing the 
column densities of HI and metals, we fit the Lyman beta spectrum to obtain 
the column densities in each HI line, then we sum over all 
components for each species which lie within the region where the
Lyman beta transmission is below 0.7.  This technique
is similar to that used
by SC, and it avoids the ambiguous and often physically meaningless task
of associating each metal line with a subcomponent in HI.
The metal line absorption in systems where 
the Lyman beta transmission does not fall below 0.7
is too weak to be detected at the SC limits.
 
The CIV/HI column density ratios for all such systems detectable in Lyman alpha
and CIV in the 400 random LOS at $z=3$ are plotted as circles in 
Figure \ref{fig2}.
Observed values from SC (with $\log{N_{\rm HI}} > 5 \times 10^{14}
$) are shown as solid triangles.  The crosses will be discussed below.
The modest scatter in CIV/HI predicted by our uniform metallicity model
arises from a scatter in physical densities and temperatures of the absorbing 
regions. Because there exists a fairly tight correlation between HI column 
density and physical density and temperature in the model (Miralda-Escud\'{e} 
\etal 1996, 1997), this scatter is small.
The scatter in the observational data is seen to be at least an order of 
magnitude higher.
We thus confirm the findings of Rauch, Haehnelt \& Steinmetz (1996) that 
significant spatial variations in metallicity are required in order to explain
the observations of CIV/HI in the high-redshift universe in the context of
this model for the Lyman alpha forest. 

The model datapoints scale directly with the carbon abundance, and with
[C/H]$\approx -2.5$ the mean value of CIV/HI of the points from the simulation
above the SC detection limit (dashed line) matches that of the observational 
datapoints. Essentially all systems with $\log{N_{\rm HI}} \simgt 15.2$ 
(the Lyman beta selected sample of SC) have
detectable CIV absorption, and these systems constitute the major fraction
of the observations. For lower values of $N_{\rm HI}$ we expect an increasing
fraction of the CIV lines to lie below the detection limit, but if the systems
with $\log{N_{\rm HI}} \approx 14.8$ have the same metallicity distribution as
the higher column density systems, the model still predicts that the  
density of observable systems around 14.8 should be several times larger
than, say, around 15.5. In the SC data, only a small cluster of systems, 
with $N_{\rm CIV}$ close to the detection limit, is found around 14.8.
This suggests that the mean metallity [C/H] of these lower column
density systems is significantly less than $-2.5$, and that we are 
only observing
the high-metallicity tail of a large population of systems with a mean
CIV/HI below the detectability limit.  Thus, comparing our model to
observations indicates that low column density forest systems
($\log{N_{\rm HI}} \simlt 15$) have a lower mean metallicity than the
higher column density systems. This conclusion is clearly insensitive to the
details of the adopted model: it holds for any model that 
predicts a roughly constant mean CIV/HI value in the interval $14.8 <
\log{N_{\rm HI}} <15.5$.

Gnedin \& Ostriker (1997), simulating the reionization and
metal enrichment of the universe by population III stars, predict (at $z=4$)
a pattern of metal enrichment that fits our requirements remarkably well.
They predict a mean metallicity of [C/H]$\sim -2.3$ with a large 
scatter, similar to what we infer by comparing our simulation to the
SC data.  They further predict
a dependence of metallicity on volume density (and hence on column density,
due to the correlation mentioned above) because most of the
star formation and subsequent enrichment of the surrounding IGM
take place within relatively dense regions,
whereas regions with low densities corresponding to column densities  
$N_{\rm HI} \simlt 10^{13.5} {\rm cm}^{-2}$ have essentially no metals.

Figure \ref{fig2} also shows a sample of high column
density systems ($\log{N_{\rm HI}}>16$) with their CIV/HI ratio shown 
as crosses. These lines of sight were pre-selected to have high total 
HI column density, in order to provide better statistics on the absorption
properties of the relatively uncommon high column density systems. 
It is seen that the model
predicts a systematically decreasing value of CIV/HI with $N_{\rm HI}$.
This trend is consistent with the observations of SC,
who find that Lyman limit systems with $17 \simlt \log{N_{\rm HI}}
\simlt 18$ have a mean log(CIV/HI)$\sim -4$ and show that the trend cannot
be explained by radiative shielding effects alone. The physical reason
for this trend in the model is the correlation between
the HI column density and the physical gas density. For a 
homogeneous radiation field, the photoionization parameter, defined as the
ratio of ionizing photons to free electrons, will be a 
decreasing function of $N_{\rm HI}$. For relatively low values of 
$N_{\rm HI}$, the CIV column density increases slightly faster than
$N_{\rm HI}$ with $N_{\rm HI}$, so the CIV/HI ratio is a weakly 
increasing function. When $N_{\rm HI}$ exceeds a value $\sim 10^{16}
{\rm cm}^{-2}$,  $N_{\rm CIV}$ starts to drop as the CIV recombines to
lower ionization states, causing the observed decrease in CIV/HI.  
Rauch et al.\ (1996) find a similar trend of CIV/HI with $N_{\rm HI}$.
For Lyman limit systems ($\log{N_{\rm HI}} \simgt 17.2$) the ionization 
parameter will decrease even faster with $N_{\rm HI}$ because the systems
become increasingly opaque to the Lyman limit photons of the background 
radiation.

\section{Redshift evolution and observed fractions of metal lines at $z=3$}
Figures 3a--3d show the magnitude and scatter of the column densities
in CIV, CII, SiIV, and NV at five different redshifts. The 
systems studied here are chosen to mimic the Lyman beta selected 
sample of SC: only systems that are Lyman beta saturated
($\log{N_{\rm HI}}>15.2$)
are included, and of those, only systems with metal column densities
larger than the detection limits of SC ($10^{12}$ for CII, CIV, NV, and
$5 \times 10^{11}$ for SiIV) are plotted.  
At $z=3$, 98 of our 400 random LOS had absorbers with $\log{N_{\rm HI}}>15.2$.
We have adjusted the metallicity of our model of this high
column density sample to [C/H]$=-2.3$, which reproduces the observed mean value 
$\log{N_{\rm CIV}}=13.2$.
As expected from the discussion of Figure 2, the scatter in column density
is less than observed, presumably because the model assumes a 
homogeneous metallicity distribution.

The model shows no significant evolution in the mean column 
densities with redshift.
This is consistent with what is seen observationally (see Figures 4a--4d of SC).
For CIV, there is a hint of evolution, with the mean value of $N_{\rm CIV}$
for $z=2$, 2.3,
and 2.7 being around $10^{13}$, while at $z=3$ and 3.5 it is around
$10^{13.2}$, but the scatter is consistent with no evolution.  For
the other species, the observational detection limits combined with
small numbers of detections make it difficult to draw conclusions regarding
the redshift evolution of their mean column densities, but again the results
are consistent with no evolution.
The model also shows, at face value, that the number of detectable metal systems
increases dramatically with redshift (the larger number of points in Figures
3a--3d with increasing $z$), an effect that is not nearly 
so pronounced in the observations.  The increase
arises partly as an artifact of our
simulation volume representing a larger redshift interval at earlier
times ($\Delta \lambda \propto (1+z)^{3/2})$, and 
partly because the universe is 
physically denser at higher redshifts, causing an increase in  the number of 
Lyman alpha and weak metal absorption lines per unit wavelength.
 
The model as well as the data show that while essentially all systems in
the Lyman beta selected sample have detectable CIV absorption, only a fraction 
of the absorbers have column densities in SiIV, CII, and NV above the detection 
limit.  
To check our model quantitatively, we 
have taken the output at $z=3$ as representative for the sample observed by SC 
and added an additional Gaussian scatter of 
$\sigma ([{\rm C}/{\rm H}]) = 0.8$ in 
metallicity so that the observed scatter in $N_{\rm CIV}$ is crudely reproduced.

The column densities without and with this added scatter are displayed as
triangles in Figures 4 and 5, respectively, and the corresponding
fractions of the systems above the detection limit (Figure 5) are shown 
in column 1 of 
Table 1. This model is seen to predict too few SiIV systems and too
many NV systems relative to what is observed (column 3 of Table 1).
No reasonable changes in the relative abundances [Si/C] and [N/C] can remove 
this discrepancy. If, however, the ionizing radiation field is softer than
the HM spectrum we will, for the relevant ionization conditions
in the Lyman alpha absorbers, expect to see less NV (more NIV) and more SiIV 
(less SiV) and CIV, while CII has an ionization potential too low to be
significantly affected. The model column densities after modifying the HM
spectrum by reducing the intensity by a factor of 10 above
4 Rydbergs are shown as circles in Figure 4 and are seen to follow the
expected trends. Most dramatically, the many low column density SiIV systems
that exist in a HM field are seen to have column densities
close to or above the detection limits after the removal of the energetic
photons. 

The predicted distributions of column densities above the detection limit
for the normal HM spectrum versus the softened background spectrum are 
plotted in 
Figure 5. Gaussian scatter of $\sigma ([{\rm C}/{\rm H}]) = 0.8$ has been added
to both distributions, and the mean [C/H] in the softened model has been 
adjusted 
to fit the observed mean value of $N_{\rm CIV}$, requiring
[C/H]$=-2.85$ (note the lower mean CIV as compared with the circles
in Figure 4). The fractions above the detection limits for the softened model 
are
shown in column 2 of Table 1. For this choice of background radiation there is 
much better 
agreement between the predicted and observed metal line fractions for
NV and SiIV.  Because of the reduction in metallicity, however, slightly
too few CII detections are now seen.
 
Also shown in Table 1 is the mean HeII optical depth in the models.
Cutting the number of HeII ionizing photons by a factor of 10 increases 
$\tau _{\rm HeII}$ from 1.6 to 3.6. Observational results by
Hogan, Anderson \& Rugers (1996) 
indicate $1.5 < \tau_{\rm HeII} <3$ (95 \% confidence level)
at $z\approx 3.3$, and at $z=3$ we would expect it to be somewhat less. 
A cut by a factor of 10 thus seems to be barely acceptable, and the spectrum
giving the best fit to the observational data is probably somewhere in between 
the two extremes considered.  Alternatively, the observations could
be accomodated if the ionizing background was inhomogeneous, and
was softer in some regions compared with others due to varying
distances from the nearest ionizing source.
Given the crudity of the approach made in 
producing Figure 5 and Table 1 and the still sparse amount of observational 
data, we will not attempt to infer any further details of the radiation
field. We simply suggest, based on the above results,  that the ionizing 
background photon field around $z=3$ is somewhat softer than the 
HM spectrum.

We now consider the SiIV/CIV ratio as a signature for HeII reionization.
SC suggest that this ratio increases rather abruptly by a factor of a few
around $z=3.2$, and that this effect could be explained if the HeII reionization
redshift is around this value. While it is true that the SiIV/CIV ratio in
a uniform density, one phase medium will increase as high energy photons are
removed from the spectrum by HeII absorption, the effect is different in our
models. As can be seen from Figure 4, the mean column density of the 
{\em observable} SiIV systems will not increase if the radiation field is 
softened, because of the numerous SiIV systems getting shifted upwards to just 
above the detection limit. Meanwhile, $N_{\rm CIV}$ is seen to increase if the 
radiation
field is softened, so the net effect is actually to {\em decrease} the SiIV/CIV
ratio. If the observed increase in SiIV/CIV is real, we are seeing the first
clear example of a failure of the model to reproduce observations.    
It should be remarked, however, that our model predicts optical depths
for HeII absorption that are consistent with observations without the need
to assume HeII reionization at $z=3.2$ (Croft \etal 1996). 

\section{Conclusions}
We have compared a cosmological simulation of the Lyman alpha
forest and associated metal lines to recent observations
of CIV, CII, SiIV, and NV absorption in high-redshift QSO spectra.
Our simulated spectra have signal-to-noise properties similar to
those of the observed spectra, and we analyze them by a Voigt-profile
fitting procedure that is similar to that used in the observational analyses.
We can examine our results from two different points of view.
First, we can assume that our theoretical model provides an
accurate description of physical conditions in the absorbing gas
and use the comparison to the observations to draw inferences about
metal enrichment of the IGM and the spectral shape of the ionizing
radiation background.  Second, we can ask whether the predicted
properties of the IGM are consistent with the metal line observations,
within the uncertainties of the observations and the theoretical modeling.
Adopting the first point of view, we conclude:
\begin{enumerate}
\item{The mean metallicity of the IGM at $z\approx 3$ corresponds to
[C/H]$\approx -2.5$.  More specifically, for the Lyman beta selected sample 
of SC ($\log{N_{\rm HI}} \simgt 15.2$), the inferred mean metallicity 
ranges from [C/H]$\approx -2.8$ to [C/H]$\approx -2.3$,
depending on the assumed spectral shape of the ionizing background.}
\item{Significant spatial variations of [C/H], with an rms
scatter of roughly one order of magnitude, are required to match
the observed scatter in CIV/HI.}
\item{There is a trend between mean enrichment and HI column density;
Lyman alpha absorbers with low HI column densities (and correspondingly
low physical densities) have systematically lower metallicities.}
\item{The ionizing background spectrum at $z \approx 3$ must be somewhat
softer than the spectrum predicted by HM, in order to explain the observed
number of SiIV systems.  However, if the spectrum is too much softer
than the HM spectrum, there will be conflict with the observed HeII
optical depth (see Croft et al.\ 1996; Davidsen, Kriss, \& Zheng 1996;
Hogan et al.\ 1996).}
\end{enumerate}

The required enrichment of the IGM (mean value, scatter, and trend
with density) is very similar to that predicted by Gnedin \& Ostriker's (1997)
numerical simulations of metal enrichment and reionization by population III
stars.  The required scatter in metallicity is also plausible on observational
grounds, since it is similar to that seen in 
halo stars in our own galaxy (e.g. Ryan \& Norris 1991). 
However, some of the observed scatter in CIV/HI could also be contributed
by inhomogeneities in the ionizing background instead of metallicity 
variations.

Within the current uncertainties of the observational data and the
theoretical modeling, we expect the predictions presented here to
be generic to cosmological models that have similar amounts of small-scale
power at $z \sim 2-3.5$.  Several results of our comparison support
the adopted physical model of the high-redshift IGM:
\begin{enumerate}
\item{With plausible assumptions about enrichment and the ionizing
background, the model reproduces the observed abundances of
CIV, CII, SiIV, and NV in Lyman alpha forest systems fairly well.
This agreement suggests that the predicted densities and temperatures of
the absorbing gas are at least approximately correct.}
\item{Consistent with SC's observations, the model predicts no substantial
evolution in the mean value of $N_{\rm CIV}$ for Lyman-beta-selected
HI lines over the redshift range $2.5 \simlt z \simlt 3.5$.}
\item{The model successfully explains the observed trend towards lower
CIV/HI ratios in high column density systems
($N_{\rm HI} \sim 10^{16}-10^{17} {\rm cm}^{-2}$).
This trend arises because of the correlation between column density
and physical density, which causes the stronger absorbers to have
more of their carbon in lower ionization states.}
\item{Qualitatively, the model reproduces observations that show
multiple CIV components associated with strongly saturated Lyman alpha lines 
(see Figure~1).  We will explore the model predictions more quantitatively
in future work.}
\end{enumerate}
The observation that is most difficult to explain in this model is the
jump in the mean SiIV/CIV ratio at $z \approx 3$, which SC suggest is
due to a rapid change in the shape of the ionizing background spectrum
associated with HeII reionization.  In our model, there are many weak
SiIV systems below the SC detection threshold.
Softening the background spectrum moves these systems above the
threshold, keeping the mean SiIV column density low.

To the extent that the analyses overlap, our conclusions agree well
with those of Haehnelt \etal (1996) and Rauch \etal (1996), despite
differences in the numerical techniques, in the mass resolution of 
the simulations, and in the type of simulations (a single large box
representing a randomly selected volume vs.\ multiple, higher resolution
simulations focused on collapsing regions).
There are many potential improvements for future modeling, including higher
resolution of the gas dynamics to better describe the structures giving
rise to the moderate column density systems where metals are detected, 
a non-equilibrium treatment of the ionization of all species, 
and explicit incorporation of spatial variations in the metallicity 
of the gas and the intensity and shape of the background radiation field.
Future studies can also examine alternative reionization scenarios
and other cosmological models, to see which theories of cosmic structure
formation can account for metal line absorption in the high-redshift 
universe.

\section*{Acknowledgements}
We have benefitted from discussions with Len Cowie, Gary Ferland,
and Bernard Pagel.  This work was supported by the NSF under grants
AST90-18256 and ASC93-18185, and the Presidential Faculty Fellows Program.
UH acknowledges support by a postdoctoral research 
grant from the Danish Natural Science Research Council. DW acknowlegdes support
from NASA grants NAG5-3111 and NAG5-3525. Computing support was provided by
the San Diego Supercomputer Center.

\clearpage
 
\begin{deluxetable}{cccc}
\footnotesize
\tablecaption{Fraction of observable metal line systems in Lyman beta sample}
\tablewidth{0pt}
\tablehead{
\colhead{Line} & \colhead{HM} & \colhead{HM+cut} & \colhead{Observed}
}
\startdata
 CII   &  24 \% & 15 \% & 24 \% \nl
 CIV   &  90 \% & 96 \% & 95 \% \nl
 SiIV  &  17 \% & 43 \% & 38 \% \nl
 NV    &  45 \% & 14 \% & 17 \% \nl
$\tau _{\rm HeII}$ & 1.5 & 3.6 & $\sim 2$
 
\enddata
\end{deluxetable}

\clearpage

\figcaption[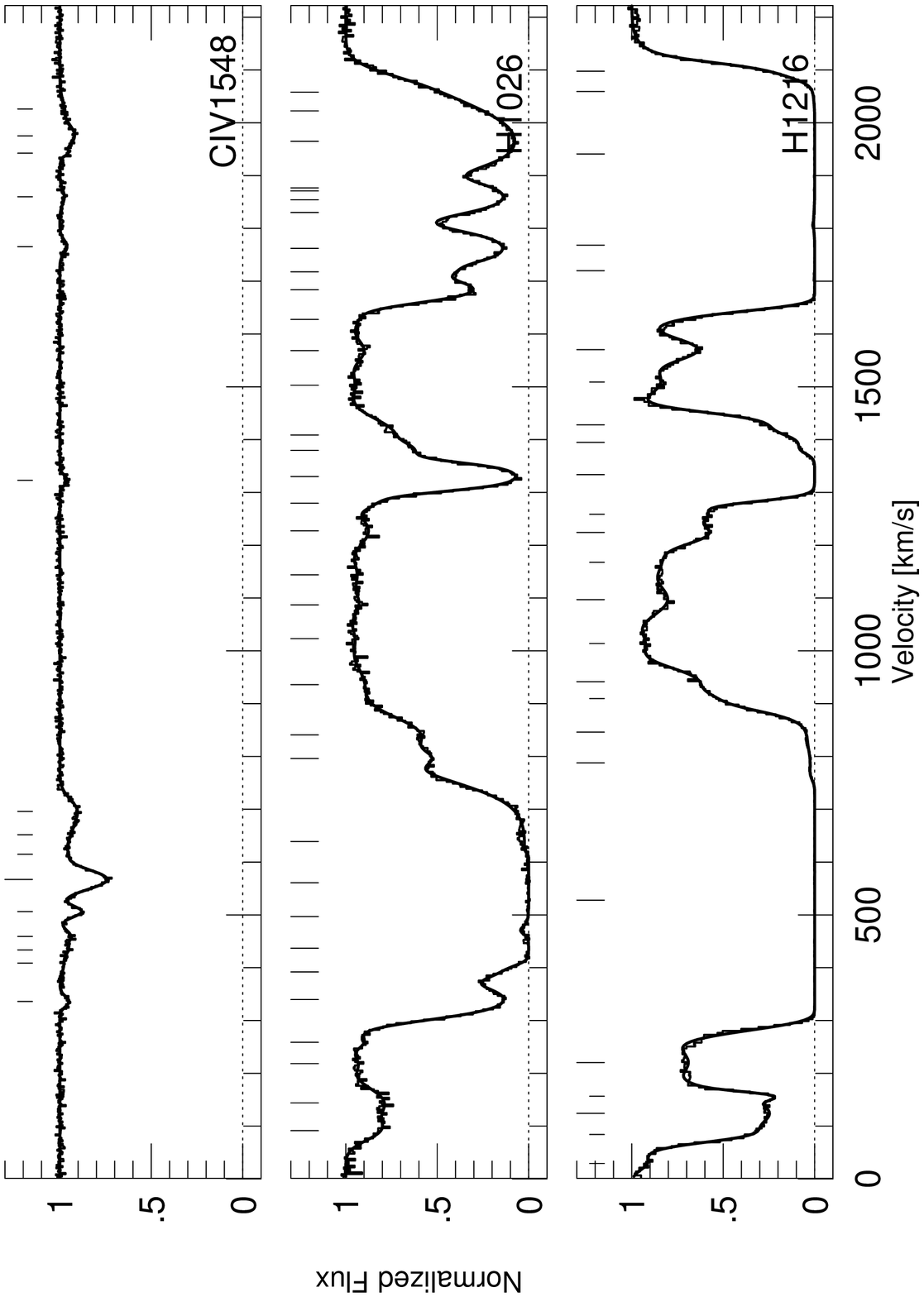]{Artificial spectra (noise added) at z=3 for HI Lyman alpha
(H1216) and corresponding Lyman beta (H1026) and CIV (CIV1548) absorption
along a line of sight with a relatively large amount of  CIV. 
The solid lines through the spectra are  Voigt profile decompositions 
obtained using AUTOVP.
Short and long vertical ticks indicate components with column density less 
than and greater than $10^{13}$, respectively.
\label{fig1}
}

\figcaption{Predicted CIV/HI ratios for the random sample (circles) and
         high column density sample (crosses) from our simulation. 
	The triangles show observed values from the spectrum of
        Q1422+231 (Songaila \& Cowie 1996), and the dashed line
	 represents a detection limit of $N_{\rm CIV}=10^{12} {\rm cm}^{-2}$.
	 The triangles directly on the line are upper limits.
The scatter in the observations is signficantly higher than that in
our uniform metallicity model.  The lack of observational detections 
of CIV/HI $ > -2.5$ for $N_{\rm HI}<10^{15} {\rm cm}^{-2}$ indicates that 
the mean metallicity in this regime is lower, or else the model would predict
an even higher density of detections than in the higher column density systems.
\label{fig2}
}

\figcaption{Metal line column densities of systems identified in 400
            random LOS at $z=2$, 2.3, 2.7, 3.0, and 3.5. Only systems
	    above the detection limits of Songaila \& Cowie (1996) are 
	    included. The errorbars next to the points indicate the
	    mean value and 1$\sigma$ scatter of the data.  There is no 
	    significant evolution in the mean column densities over 
            this redshift range.
\label{fig 3}
}

\figcaption{Column densities of CIV, CII, SiIV, and NV at $z=3$ for a 
Haardt \& Madau (1996) radiation spectrum (left columns, triangles) and a 
spectrum softened by cutting the intensity above 4~Ryd by a factor of 10 
(right columns, circles).
            Extremely weak lines below the detection limit have been included here
 	    to clarify the effects of softening the radiation field.
	    Detection limits inferred by SC are indicated as horizontal
	    lines in the figure.
\label{fig 4}
}

\figcaption{Column densities of metal line systems in the model with HM 
	spectrum (triangles) and
	    softened spectrum (circles) after adding an additional gaussian
	    scatter of $\sigma({\rm [C/H]})=0.8$ and adjusting the mean 
	    metallicity to reproduce the
	    observed mean CIV column density of $10^{13.2}$. 
            Only systems with column densities
	    above the SC detection limits are plotted.  Notice the increase
in detections of SiIV, and the decrease in detections of NV due to softening
the spectrum.  The resulting
statistics (see Table 1) agree better with observed values.
\label{fig 5}
}
\clearpage

\plotone{fig1.ps}

\clearpage

\plotone{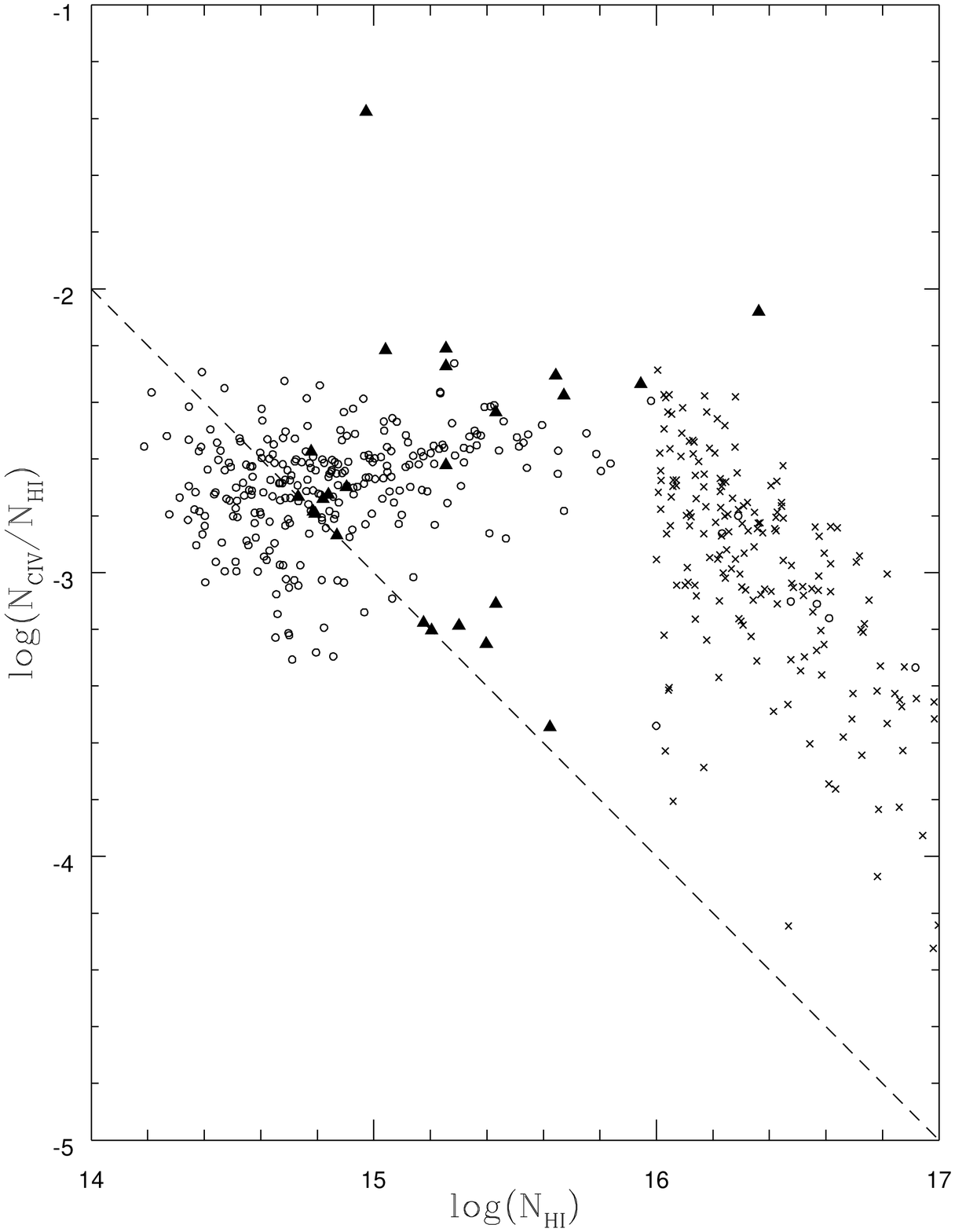}

\clearpage

\plotone{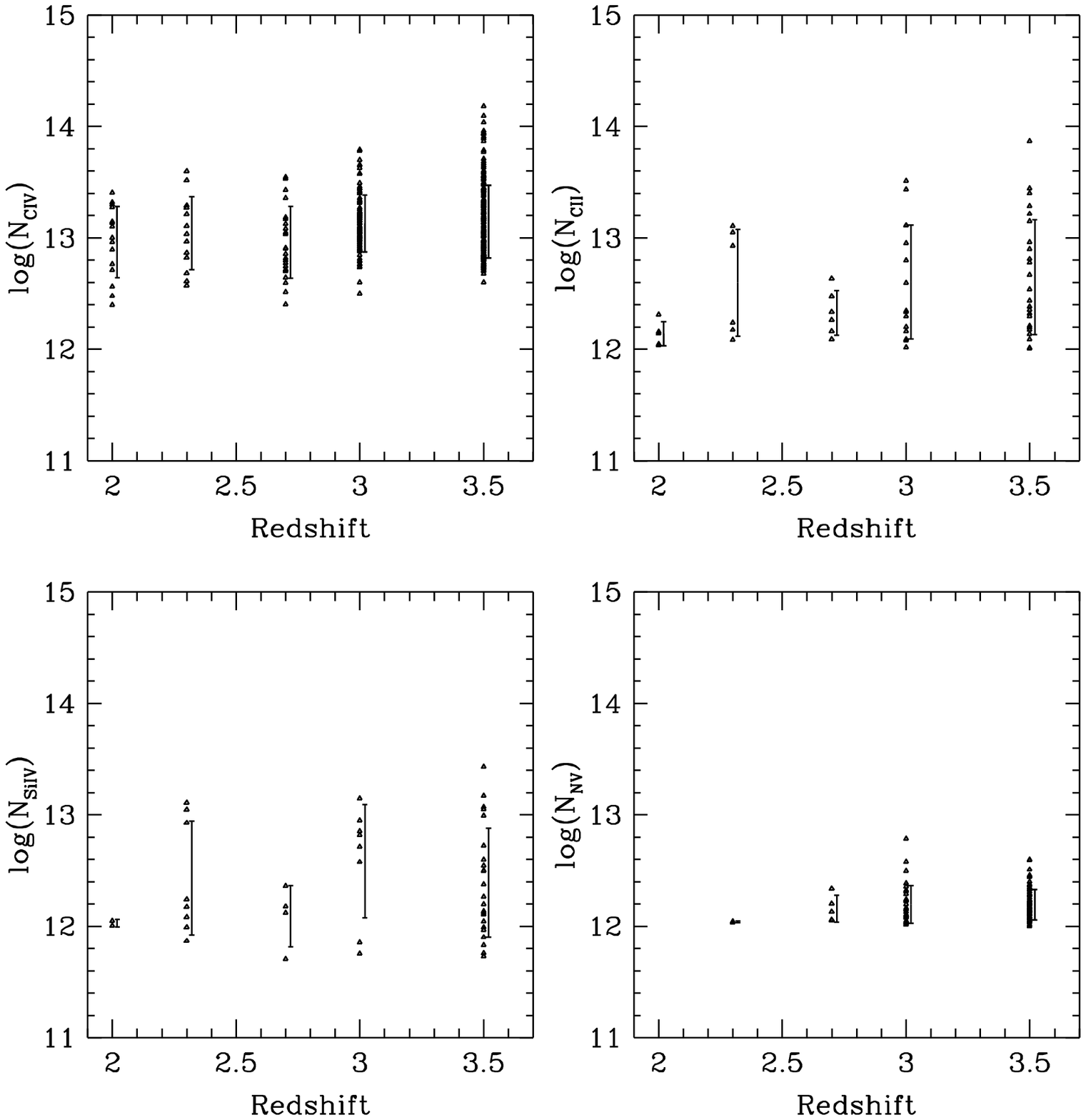}
 
\clearpage

\plotone{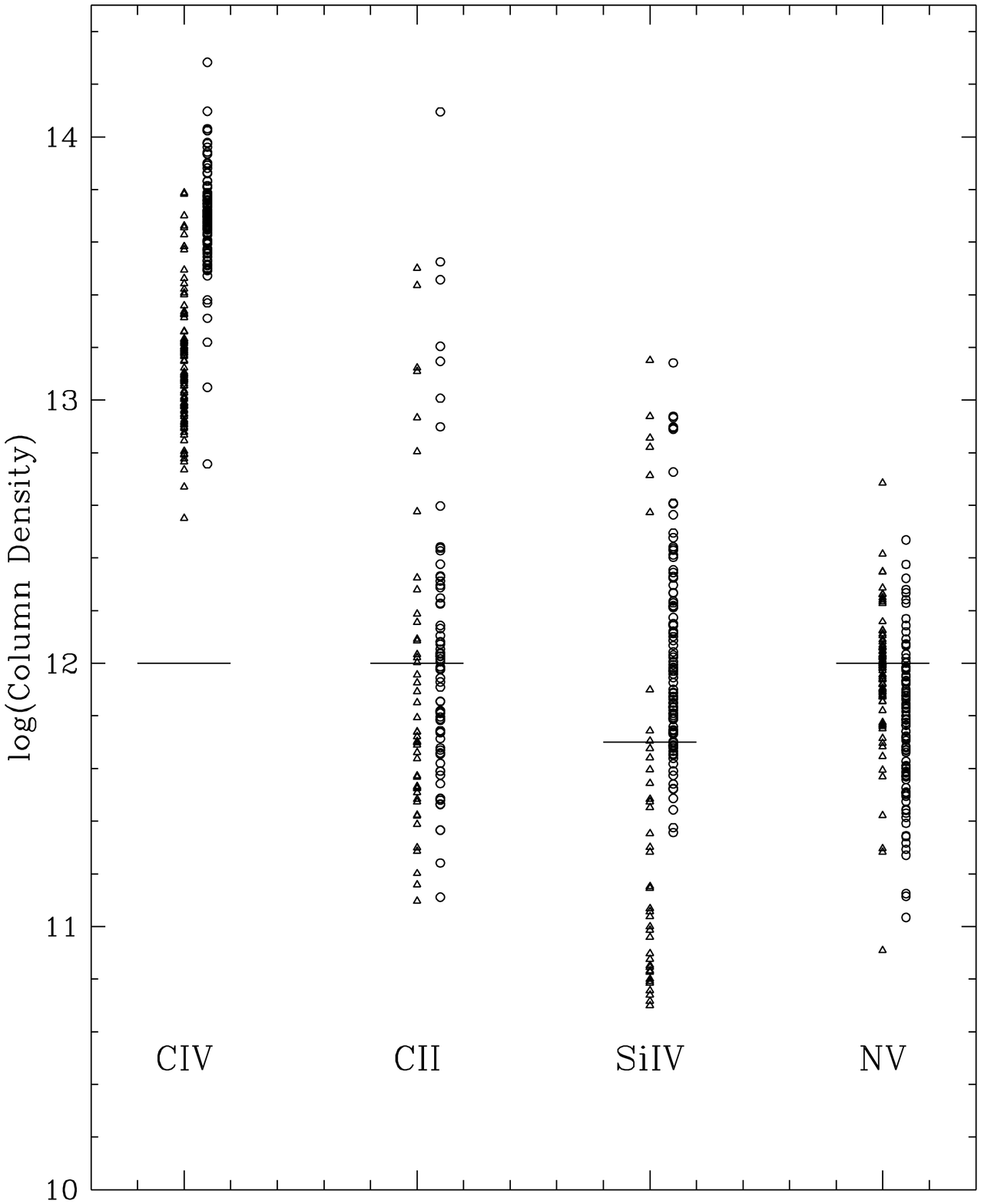}

\clearpage

\plotone{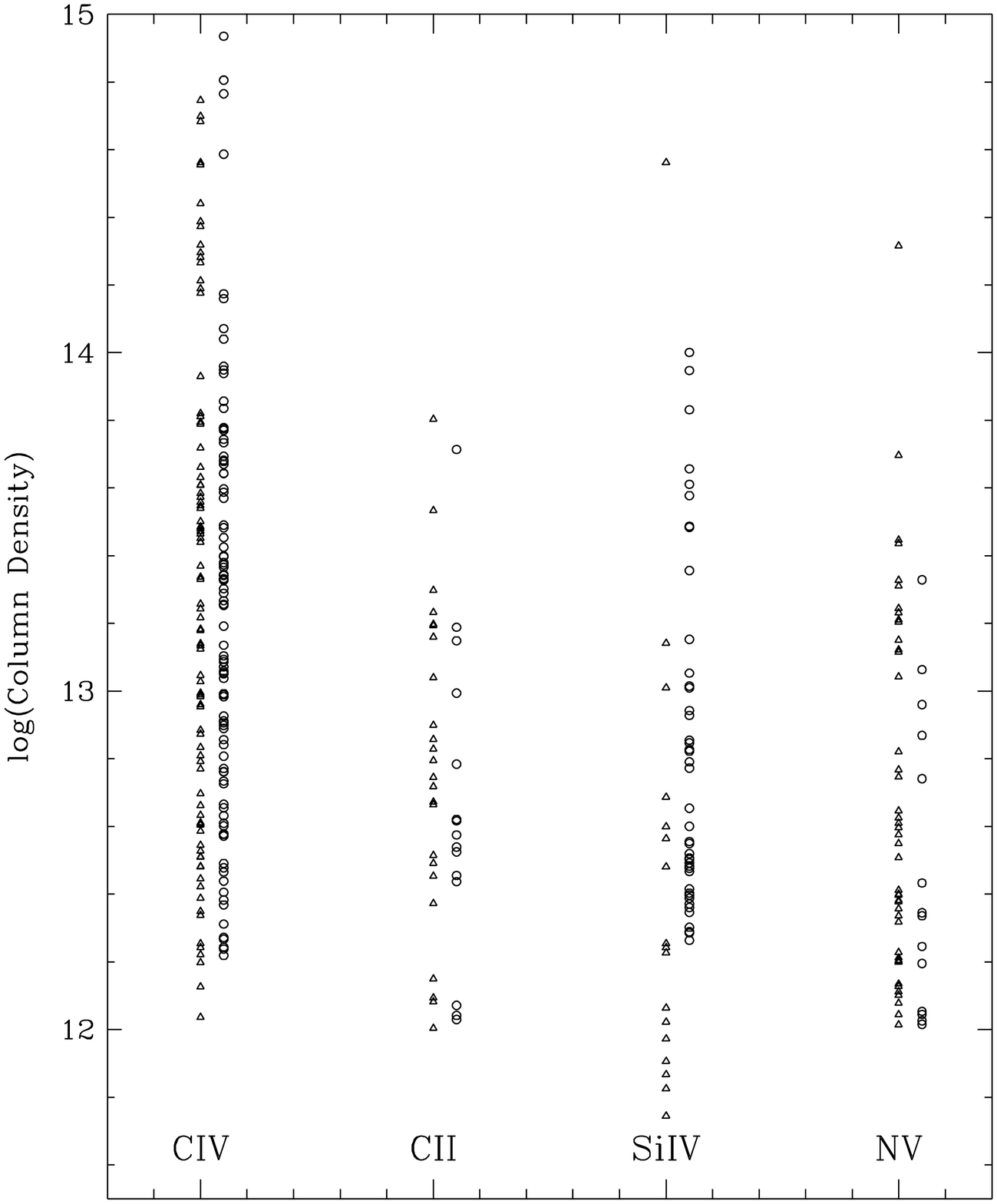}

\vfill\eject


\clearpage

\begin{thebibliography}{}
\bibitem[]{} Bi, H.G. 1993,
	\apj, 405, 479
\bibitem[]{} Bi, H.G. \& Davidsen, A.F. 1997,
	\apj, in press
\bibitem[Cen et al. 1994]{cen94} Cen, R., Miralda-Escud\'e, J.,
    Ostriker, J.P., \& Rauch M. 1994, \apj, 427, L9
\bibitem[]{} Cowie, L.L., Songaila, A., Kim, T.-S., \& Hu, E.M. 1995,
    AJ., 109, 1522 
\bibitem[]{} Croft, R.A.C., Weinberg, D.H., Katz, N., \& Hernquist, L. 1996,
   \apj, submitted
\bibitem[]{} Dav\'{e}, R., Hernquist, L., Weinberg, D.H., \& Katz, N. 1997,
    \apj, in press [DHWK]
\bibitem[]{} Davidsen, A.F., Kriss, G. A. \& Zheng, W., Nature, 380, 47 
\bibitem[]{} Ferland, G.J., 1996, University of Kentucky, Department of
    Astronomy, Internal report
\bibitem[]{} Gnedin, N.Y. \& Ostriker, J.P. 1997,
	\apj, submitted. Astro-ph/9612127.
\bibitem[]{} Gunn, J.E. \& Peterson, B. A. 1965, \apj, 142, 1633
\bibitem[HM]{haa96} Haardt, F. \& Madau, P. 1996, \apj, 461, 20 [HM]
\bibitem[]{} Haehnelt, M.G., Steinmetz, M., \& Rauch, M. 1996, \apj, 465, L65  
\bibitem[Hernquist \& Katz 1989]{her89} Hernquist, L. \& 
    Katz, N. 1989, \apjs, 70, 419
\bibitem[HKWM]{her96} Hernquist, L., Katz, N., Weinberg, D.H., 
    \& Miralda-Escud\'e, J. 1996, \apjl, 457, L51 
\bibitem[]{} Hogan, C.J., Anderson, S.F., \& Rugers, M.H. 1996,
    astro-ph/9609136
\bibitem[]{}  Hui, L., Gnedin, N.Y., \& Zhang, Y. 1997,
	astro-ph/9608157
\bibitem[]{} Katz, N., Weinberg, D.H., \& Hernquist, L. 1996,
    ApJS, 105, 19
\bibitem[]{} Lynds, R. 1971,
	\apj, 164, L73
\bibitem[]{} Meyer, D.M. \& York, D.G. 1987,
	     \apj, 315, L5
\bibitem[]{} Miralda-Escud\'{e}, J., Cen, R., Ostriker, J.P. \& Rauch, M. 1996,
	     \apj, 471, 582
\bibitem[]{} Miralda-Escud\'{e}, J., Weinberg, D.H., Hernquist, L.
 		\& Katz, N. 1997, in preparation
\bibitem[]{} Pettini, M., Lipman, K., \& Hunstead, R.W. 1995,
    \apj, 451, 100 
\bibitem[]{} Petitjean, P., M\"{u}cket, J.P. \& Kates, R.E. 1995,
	A\&A, 295, L9	
\bibitem[PRS]{pre93} Press, W.H., Rybicki, G.B., \&
    Schneider, D.P. 1993, \apj, 414, 64 
\bibitem[]{} Rauch, M., Haehnelt, M.G., \& Steinmetz, M. 1996, \apj, in press,
    astro-ph/9609083
\bibitem[]{} Ryan, S.G. \& Norris, J.E. 1991,
	AJ, 101, 5
\bibitem[]{} Sargent, W.L.W., Young, P.J., Boksenberg, A., \& Tytler, D. 1980,
    ApJS, 42, 41
\bibitem[]{} Songaila, A. \& Cowie, L. L. 1996,
    AJ, 112, 335 [SC]
\bibitem[]{}Tomkin, J., Lemke, M., Lambert, D.L., \& Sneden, C. 1992,
    AJ, 451, 100
\bibitem[]{} Wheeler, J.C., Sneden, C., \& Truran, J.W. 1989,
    ARA \& A, 27, 279
\bibitem[]{} Womble, D.S., Sargent, W.L.W., \& Lyons, R.S. 1995, 
    \apj, submitted, astro-ph/9511035
\bibitem[Zhang, Anninos, \& Norman 1995]{zha95} Zhang, Y., Anninos, P.,
    \& Norman, M.L. 1995, \apjl, 453, L57
\bibitem[]{} Zhang, Y., Anninos, P., Norman, M.L., \& Meiksin, A. 1997,
	\apj, submitted
\end{thebibliography}
\end{document}